\documentclass[aps,pre,10pt,showkeys]{revtex4-1}
  
\usepackage{url,hyperref,lineno,microtype}
\usepackage{amsmath,amsfonts}
\usepackage{color,physics,graphicx,bm,lipsum}
\usepackage{ulem}

\newcommand{\ee}{\mathrm{e}}

\begin{document}

\title[Animal encounter and interaction processes]{Misconceptions about quantifying animal encounter and interaction processes}

\author{Debraj Das}
\email{debrajdasphys@gmail.com}
\affiliation{Quantitative Life Sciences, ICTP -- The Abdus Salam International Centre for Theoretical Physics, Trieste 34151, Italy}

\author{V. M. Kenkre}
\affiliation{Department of Physics and Astronomy, University of New Mexico,  Albuquerque, 87106, NM, USA}

\author{Ran Nathan}
\affiliation{Movement Ecology Laboratory, Department of Ecology, Evolution and Behavior, Alexander Silberman Institute of Life Sciences, Faculty of Science, The Hebrew University of Jerusalem, Jerusalem 91904, Israel}

\author{Luca Giuggioli}
\email{luca.giuggioli@bristol.ac.uk }
\affiliation{School of Engineering Mathematics and Technology, University of Bristol, Bristol BS8 1UB, United Kingdom}

\begin{abstract}
Quantifying animal interactions is crucial for understanding various ecological processes, including social community structures, predator-prey dynamics, spreading of pathogens and information. Despite the ubiquity of interaction processes among animals and the advancements in tracking technologies enabling simultaneous monitoring of multiple individuals, a common theoretical framework to analyse movement data is still lacking. The diverse mechanisms governing how organisms perceive the proximity of others have led to species-specific theoretical approaches, hindering a common currency with which to evaluate and compare findings across taxa. We propose a general framework, borrowing tools from statistical physics, specifically from the theory of reaction diffusion processes. While some of these tools have been employed to predict pathogen transmission events, they have not yet pervaded the movement ecology literature. Using both continuous and discrete variables, we demonstrate the suitability of our framework to study interaction processes. Defining interactions as the transfer of information between individuals, we show that the probability of information transfer for the first time is equivalent to the first-passage probability of reacting in a multi-target environment. As interaction events reduce to encounter events for perfectly efficient information transfer, we compare our formalism to a recent approach that takes the joint occupation probability of two animals over a region of interaction as a measure of the encounter probability, rather than the first-encounter probability. We show the discrepancy between the two approaches by comparing analytically their predictions with continuous variables, while with discrete variables we quantify their difference over time. We conclude by pointing to some of the open problems that reaction diffusion formalism might be able to tackle.
\end{abstract}

\keywords{animal interactions, encounter problem, movement ecology, random walks and Brownian motion, reaction diffusion processes} 

\maketitle

\section{Introduction}
A vast number of studies in animal ecology aims at linking behaviour at the level of the individuals to the processes governing the dynamics of a group or an entire population \cite{levin1992}. Underlying this fundamental tenet is the search for general laws that link the interactions between animals to the patterns that emerge at much larger scales. A renewed interest in such perspective has surfaced in the last decades following the introduction of the movement ecology framework \cite{nathanetal2008} and the advances in sensor technologies that allow to track animals in space and time at unprecedented resolution \cite{nathanetal2022,matleyetal2022}. It is the ability to follow simultaneously multiple individuals and infer when and how they interact that will be instrumental to the understanding of this micro-to-macro connection. Notably, despite the pervasiveness of interaction processes between moving organisms, models in the animal ecology literature of how individuals interact or more simply when and where they encounter or are in proximity of one another have been limited.

While theoretical approaches that aim at quantifying interaction processes have appeared \cite{hutchinsonwaser2007,bartumeusetal2008,mckenzieetal2009,gurarieovaskainen2013,martinez-garcia_how_2020}, efforts to develop a general framework have been stymied by two main factors: semantic issues, due in part to the different ways in which animals may interact, and the apparent absence of analytical `null' models in the movement ecology literature. As animals interact by relying on their sensory biases, by using their cognitive mechanisms and by exploiting their motor abilities, finding a general definition of interaction has been challenging and the rationale has often resulted in specific choices based upon the biological questions and the species under observation.

In collective movement studies a classical example is the use of delays in motor response to determine leadership roles. This approach has been employed to
construct social ranks in a flock of pigeons based on their global delayed response in following each other's trajectories \cite{nagyetal2010}, and to extract time-dependent delays during coordinated manoeuvres of foraging bat pairs to identify leaders \cite{giuggiolietal2015} or to classify the influential neighbours during collective turns of a shoal of laboratory fish \cite{jiangetal2017}. Examples in animal social studies also abound \cite{farineetal2015}. In that context a social interaction network for a group of individuals is constructed based on the occurrence per sampling period of well defined events \cite{whiteheadbook2008}, e.g. grooming, or parent and offspring associations, and has been used to predict how processes such an infection or some form of information is spread over the network. While these and other approaches provide practical tools to estimate specific forms of relatedness, they often lack a common currency with which to make comparison across species.

Even in the simplest scenario in which an interaction is defined as an encounter, i.e. a co-location or a proximity event, model estimations differ greatly depending on how the movement is represented. The assumption that animals move deterministically, i.e. perform ballistic motion, has led to the so-called ideal gas model prediction of an exponential time dependence in the encounter probability with mean $\pi/(8 d b v)$ \cite{hutchinsonwaser2007}, whereby in a population of density $d$ a focal individual moves with constant speed $v$ and encounters other individuals when within a distance $b$. The cornerstone of the ideal gas model is the law of mass action. It posits that encounters are directly proportional to the concentration of individuals and neglects any dependence on the statistical properties of the trajectories of the moving entities. In essence it is a mean field approximation and deviates further from the actual predictions the more winding are the movement paths and the lower the density of individuals.

In the extreme limit of very diluted systems, e.g. one randomly moving organism searching for static targets, a large literature on random biological encounters have emerged in the last twenty years. The focus of that literature has been the study of target encounter efficiency when an animal's straight movement paths follow a power law distance function as compared to a sharply decaying one \cite{viswanathanetalbook2011,reynolds2015}, the latter characteristics of Brownian motion. Various scenarios have been considered including the distinction between destructive searches, for which a target is consumed upon encounter, and non-destructive searches, for which the target is uninfluenced by the searcher \cite{santosetal2004,bartumeusetal2005}, as well as the difference between hard encounters, which occur whenever a searcher is within a threshold distance from a target, and soft encounters whose occurrence depends on some smooth functional dependence of the searcher-target distance. Given the vast number of animal interactions an important study that has brought clarity to the literature is the one by Gurarie and Ovaskainen \cite{gurarieovaskainen2013}, which has provided a classification of the different types of animal encounter interactions and has reviewed and compared many of the theoretical results, in particular of interest to us here, the findings on what is generally referred to as random search statistics (see e.g. \cite{bartumeusetal2014,bartumeusetal2016}).

In comparison to the vast literature on search of static targets, past ecological investigations on moving targets, that is on actual animal encounters, have been limited, with the exceptions of a couple of analytic studies in one dimension \cite{tejedoretal2011,giuggiolietal2013}, and two-dimensional simulation studies on animal encounters when moving as L\'{e}vy walkers both in terms of encounter efficiency \cite{bartumeusetal2008} and in terms of survival advantage when the energy content of the prey is accounted for \cite{faustinoetal2007,faustinoetal2012}. Lately, following the improved resolution in tracking technology \cite{nathanetal2022}, there has been an upsurge of interest on encounter processes  \cite{martinez-garcia_how_2020,noonanetal2021,alberyetal2021,yangetal2023,noonanetal2023}. Yet, the animal ecology literature seems to have missed out a body of work in statistical physics on the theoretical investigations of encounter and transmission events, normally referred to as the theory of reaction diffusion processes. That theory was laid out in the '80s by Kenkre in the context of exciton annihilation in molecular crystals as well as in the general field of exciton capture in sensitized luminescence \cite{kenkre1980,kenkre1982,kenkre_master_1982}. Originally the theoretical formalism was developed for movement in unbounded discrete lattices with focus on coherence in exciton motion \cite{kenkrewong1981,kenkre_master_1982}, but specific problems were also solved in bounded systems \cite{kenkrewong1980}. A focused aim of those investigations was the resolution of annoying paradoxes that had been encountered in the field of molecular crystals regarding both magnitude and temperature dependence of exciton diffusion constants extracted from experimental data in aromatic hydrocarbon crystals \cite{kenkreschmid1983,kenkreetal1985}. A decisive demonstration of the errors made in previous analyses in molecular crystals was given by Kenkre and Schmid in the papers referenced. This was done in the context of the extraction of motion parameters from mutual annihilation observations on the one hand and sensitized luminescence observations on the other. A study of that demonstration would be highly useful in any encounter context whether molecular or ecological.

The techniques used to interpret empirical observations on exciton annihilation have actually been extended to spatially continuous domains to study hard encounters in an ecological context, more precisely to predict the probability of interaction for animals living within separate home ranges in one \cite{kenkresugaya2014} and two dimensions \cite{sugayakenkre2018}. By representing the tendency of an animal to remain close to its burrow or nest via an Ornstein-Uhlenbeck process \cite{giuggiolietal2006,giuggiolikenkre2014}, i.e. by tethering its motion using a spring force, an exact analytic representation of the encounter and transmission probability when interaction occurs within a cut-off distance has been derived \cite{kenkresugaya2014,sugayakenkre2018}. The formalism that Kenkre developed with Sugaya in this context towards the implementation of the reaction diffusion theory has been given in detail in the recent publication of a book by two of the co-authors (see chapter 6 in Ref. \cite{kenkregiuggiolibook2021}). Even though these analytical techniques clearly represent the most appropriate and powerful starting point from which to study a broad range of encounter and interaction processes, surprisingly they have not been exploited in the animal ecology literature.

Following Kenkre's reaction diffusion approach, a novel analytic formalism to study movement on discrete lattices and in discrete time \cite{giuggioli_exact_2020} has allowed to derive analytically the so-called splitting probabilities, that is the probability for interaction events to occur in a set of locations and not others \cite{giuggiolisarvaharman2022}. Knowledge of these splitting probabilities allows to predict interactions in a multi-target environment, and has lead to analytic predictions of the spatio-temporal dynamics of random transmission events in arbitrary dimensions and arbitrary (lattice) topology \cite{giuggiolisarvaharman2022}, including hexagonal and honeycomb lattices \cite{marrisetal2023}, as well as when individuals undergo a resetting dynamics \cite{dasgiuggioli2022} or when the environment is spatially heterogeneous \cite{sarvaharmangiuggioli2023}. All these developments, both with continuous and discrete variables, should form the backbone of a general theory of animal interaction and encounter processes, and given their analytic formulation, should become part of the arsenal of `null' models in movement ecology.

Here we present evidence of the need of a reaction diffusion formalism to study encounter and transmission events between animals, interchangeably referred to as walkers in this study. We define a
transmission event as the first occurrence when information is successfully transferred between two individuals. With continuous variables we consider the spatio-temporal dynamics of two diffusing animals (Brownian walkers) living in two separate home ranges undergoing Ornstein-Uhlenbeck motion and show the analytic formalism that has been developed in that case \cite{sugayakenkre2018} to represent the probability of first-transmission.
With perfect efficiency of information transfer the first-transmission event reduces to a first-encounter event, hence aligning our definition of an encounter event to that of a first-hitting event that has been used in the ecological literature \cite{gurarieovaskainen2011}. In this limit we compare the formalism to the one presented in a recent theoretical investigation by Martinez-Garcia and collaborators \cite{martinez-garcia_how_2020} where a pair-wise distance threshold probability has been proposed as a tool to study animal encounters. For the Ornstein-Uhlenbeck case considered, we derive analytically the mathematical equation that relates the two probabilities.

For the discrete space-time formulation we also consider two diffusing animals (lattice random walkers) living in separate home ranges. We choose two scenarios to represent the characteristic reduction in movement range. In the first one, we impose a hard constraint on the movement range of the animals (reflected lattice walkers). In the second one, we account for the animal tendency to return to a den or a burrow by resetting its location at random times to its own focal point in space (resetting random walkers). For these two cases for simplicity we restrict the interactions to when animals are co-located on the same site and we quantify the time-dependence of the first-encounter probability (maximal information transfer efficiency). We compare this dependence to the one obtained by spatially integrating the animal joint occupation probability of all possible interaction co-locations, a quantity analogous to the pair-wise distance threshold probability examined with continuous space-time variables. For the discrete case we also show the exact formalism to extract mean first-transmission times.

In the present study we make various assumptions about the animals' behaviour, their environment and how we characterise their movement. In choosing very simple representations of how animals move within a home range in one and two dimensions, we have purposely sacrificed ecological complexity to gain in conceptual and mathematical transparency. We have disregarded that animals may engage in activities other than foraging (see e.g. examples in Refs. \cite{moralesetal2004,ovaskainenetal2008,gurarieetal2009} and for relevant techniques developed to infer behavioural shifts from tracking data). We have also assumed that animals move in a homogeneous environment and have represented in a simple manner how the presence of a home range in one and two dimensions affect their motion. A third assumption is that we have neglected correlations in the movement steps, which means that when animals move with some degree of persistence, our estimation of encounter and transmission rates are valid for time scales longer than the correlation or persistence time \cite{gurarieetal2009}.

\section{Materials and Methods}
\subsection{The continuous space-time formalism}
\label{sec:gen_inter}

\begin{figure}[!htbp]
\centering
\includegraphics[scale=1]{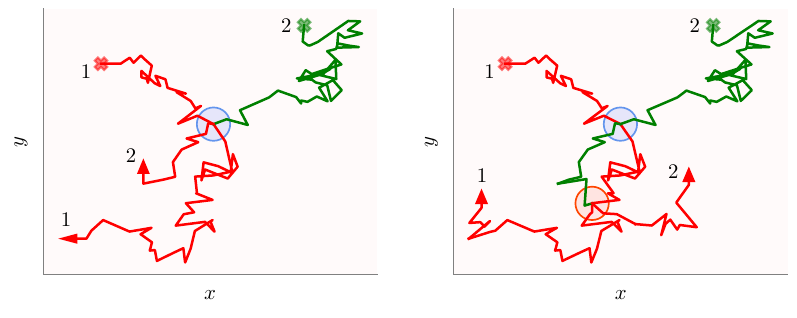}
\caption{Schematics of the two-dimensional movement paths of two animals tracked over a certain period of time that may transfer information when they are within a certain distance threshold. The circular disks represent all the spatial locations when the two individuals are simultaneously within a threshold distance from each other. Animal 1 (red trajectory) carries information, while animal 2 (green trajectory) initially does not. Both the walkers start from their respective initial points shown as the cross marks, and when information is transferred from the first to the second animal the green trajectory turns red. In the left panel the information transfer process occurs early on (blue disk), that is on the first occasion in which they are within interaction distance, while in the right panel, transfer occurs on the second occasion (red disk). The inefficiency of the transfer process is evident in both panels because the green trajectory does not turn red when on the disk boundaries (first-encounter), but only after some time the animals are within the disk. Note that time stamps of the trajectory are not explicitly indicated and the animal paths should not be thought of representing movement with constant speed. In other words the disks aim to display direct interactions, i.e. when individuals are within a threshold distance at the same time, rather than indirect ones when spatial coincidence may occur at different times.}
\label{fig:schematic}
\end{figure}

To bypass any potential semantic issues, we restrict our study and define an interaction when a measurable unit of information is being passed from one individual to the other. Examples include an infectious disease, which is transmitted through the transfer of a pathogen, or the passing of knowledge, e.g. food source location. In these cases, when the movement statistics is Markov and the information being transferred is a binary variable (presence/absence), transmission events can be modelled as a first-absorption process \cite{spendierkenkre2013}. In other words by defining interactions based on the transfer of a token of information from one animal to another, it is possible to model mathematically interaction events as a function of the movement statistics and the ability of the uninformed individual to receive information from the informed one, as exemplified pictorially in Fig. \ref{fig:schematic}. Modelling and quantifying such events and identifying the underlying principles under which randomly moving particles or more generally biological agents react with each other is an important area of investigations in statistical physics and is referred to as the theory of reaction diffusion processes.

A well-known assumption to estimate interaction times consists of summing the average time for two individuals to be in proximity, $T$, and the average reaction time or information transfer time upon proximity, $I$. Such assumption goes under different names in different disciplines, e.g. the inverse addition law in chemical reactions \cite{soustellebook2011}, or Matthiessen's rule in solid state physics \cite{zimanbook2001}. Kenkre and his collaborators showed the limitations of such an assumption \cite{kenkreetal1985}, whose validity is restricted to the so-called reaction limited case ($T/I\rightarrow 0$) and the motion limited case ($I/T\rightarrow 0$), and developed an analytic formalism to predict the time-dependent first-transmission probability between randomly moving entities in unbounded lattices \cite{kenkre1980,kenkre1982,kenkre_master_1982}. With a similar theoretical construct it is possible to analyse the transmission problem of two animals, one informed and the other one uninformed, living in separate home ranges. By representing them as two Brownian walkers biased towards their respective focal points in space, i.e. their home range centres, through a spring force (Ornstein-Uhlenbeck process), Kenkre and Sugaya \cite{kenkresugaya2014,sugayakenkre2018} have derived analytically the transmission probability, that is the probability that the uninformed (susceptible) individual has become informed (infected) at time $t$.

To understand what are the key ingredients necessary to quantify reaction diffusion processes, in particular the time-dependent transmission probability of a token of information from one individual to another, we report here some of the necessary mathematical details with continuous variables. 
We start by considering the partial differential equation (PDE) governing the dynamics of the joint occupation probability of the two tethered Brownian walkers, one susceptible and one infected, subject to an interaction rate $\mathcal{C}$ upon proximity \cite{kenkresugaya2014,sugayakenkre2018}. The PDE of the time-dependent joint occupation probability of walker 1 and 2 to be at $\vb*{r}_1$ and $\vb*{r}_2$, respectively, contains a Smoluchowski term to describe the movement to which a transmission interaction term in the form of a loss is added \cite{kenkresugaya2014,sugayakenkre2018}. A variable transformation from the coordinates $\bm{r}_{1,2}$ of the two animals $\bm{r}_+=\bm{r}_1+\bm{r}_2$ (a centre of mass coordinate would be $\bm{r}_+/2$) and a relative position $\bm{r}_-=\bm{r}_1-\bm{r}_2$ (in Ref.~\cite{sugayakenkre2018} $\bm{r}_{\pm}$ are defined with a multiplicative factor $2^{-1/2}$) allows one to write the joint PDE governing equation as~\cite{sugayakenkre2018}
\begin{align}
\pdv{P(\vb*{r}_+,\vb*{r}_{-},t)}{t} & = \grad_+ \cdot \qty[ \gamma (\vb*{r}_+ - \vb*{R}_+) P(\vb*{r}_+,\vb*{r}_{-},t) ] + \grad_{-} \cdot \qty[ \gamma (\vb*{r}_{-} - \vb*{R}_{-}) P(\vb*{r}_+,\vb*{r}_{-},t) ] 
  \nonumber \\
& + 2 D \qty( \laplacian_+ + \laplacian_{-}) P(\vb*{r}_+,\vb*{r}_{-},t)- {\mathcal{C}} \int' \dd{\vb*{r}'_+} \dd{\vb*{r}'_{-}} \delta(\vb*{r}_+ - \vb*{r}'_+) \delta(\vb*{r}_{-} - \vb*{r}'_{-}) P(\vb*{r}_+,\vb*{r}_{-},t)  \, ,
\label{eq:2walkOU}
\end{align}
where $D$ is the diffusion constant of both animals,  $\grad_{\pm}$ represents the partial differential operator in radial coordinates for $\vb*{r}_{\pm}$, $\delta(z)$ is the Dirac delta function, $\vb*{R}_{\pm}= \vb*{R}_1 \pm \vb*{R}_2$ are the transformed coordinates of the two animals' home range centres, $\gamma$ is the strength of the attraction (spring force constant) towards their respective home range centres, and the prime symbol of the integral means that integration is over a given range of values to be specified.

Note that $P(\vb*{r}_+,\vb*{r}_{-},t)$ in Eq. (\ref{eq:2walkOU}), which describes the dynamics in a 4-dimensional space, is the spatio-temporal dependence of the joint occupation probability (distribution) of the informed and uninformed individual. When a transmission event occurs, the uninformed individual disappears, and thus the probability $P(\vb*{r}_+,\vb*{r}_{-},t)$ is identically zero. This aspect is captured by the last term of Eq. (\ref{eq:2walkOU}), which indicates that there is a probability loss over time at rate $\mathcal{C}$ when the two animals are within the interaction distance, indicated by the prime integration with respect to the separation distance variable. When the rate $\mathcal{C}=0$, there is no interaction, while an encounter event is represented with $\mathcal{C}\rightarrow \infty$. Note also that the presence of the integration allows to specify the spatial locations where interactions may occur (integration over $\vb*{r}_+$) and at what distance it may occur (integration over the relative coordinate, $\vb*{r}_-$).

With $P(\vb*{r}_+,\vb*{r}_{-},t)$ non-zero when both the informed and the uninformed individuals are present, the first-transmission probability, $\mathcal{T}_{\vb*{r}^0_{\pm}}(t)$, that is the probability (density) that a first-transmission event has occurred anywhere in the interaction region is simply given by
\begin{equation}
\mathcal{T}_{\vb*{r}^0_{\pm}}(t)=\mathcal{C}\int \dd{ \vb*{r}_+} \int^{\prime} \dd{\vb*{r}_-} P(\vb*{r}_+,\vb*{r}^0_+,\vb*{r}_{-},\vb*{r}^0_-,t),
\label{eq:inf_prob_t}
\end{equation}
where the symbols $\vb*{r}^0_{\pm}$ indicate the dependence on the initial conditions $(\vb*{r}^0_+,\vb*{r}^0_{-})$ via $P(\vb*{r}_+,\vb*{r}^0_+,\vb*{r}_{-},\vb*{r}^0_-,t)$, which represents the solution of Eq. (\ref{eq:2walkOU}) the two animals are localised at $\vb*{r}^0_{\pm}$ at time $t=0$.
In Eq. (\ref{eq:inf_prob_t}) we have dropped the prime superscript on the integration over $\vb*{r}_+$, since we consider it over the entire two-dimensional space, while we have kept it for the relative coordinates since that is only over the interaction region.

To proceed further one needs to find the solution $P(\vb*{r}_+,\vb*{r}^0_+,\vb*{r}_{-},\vb*{r}^0_-,t)$ of Eq. (\ref{eq:2walkOU}) and then insert it into Eq. (\ref{eq:inf_prob_t}) to obtain the first-transmission probability. In some situations, like the one we are analysing here, the solution can be found analytically in terms of quantities that can be derived from the dynamics in the absence of interactions ($\mathcal{C}=0$). This is accomplished by employing the so-called Montroll's defect technique \cite{montroll_effect_1955,kenkrebook2021}, which allows to find analytically the Laplace transformed $\widetilde{P}(\vb*{r}_+,\vb*{r}^0_+,\vb*{r}_{-},\vb*{r}^0_-,\epsilon)$---$\widetilde{f}(\epsilon)$ represents the Laplace transform of a function $f(t)$, i.e. $\widetilde{f}(\epsilon)=\int_{0}^{\infty} \dd{t} f(t) \ee^{-\epsilon t}$, $\epsilon$ being the Laplace variable. More precisely one may express the first-transmission probability analytically as a ratio of quantities in Laplace domain defined independently of the transmission phenomenon, namely \cite{sugayakenkre2018}
\begin{equation}
\widetilde{\mathcal{T}}_{\vb*{r}^0_{\pm}}(\epsilon)=\frac{\widetilde{\mu}(\epsilon)}{\frac{1}{\mathcal{C}}+\widetilde{\nu}(\epsilon)},
\label{eq:inf_prob_lap}
\end{equation}
whose time dependence can be found numerically by performing an inverse Laplace transform. In Eq. (\ref{eq:inf_prob_lap}), the quantity $\mu(t)$ represents the probability, in the absence of any interaction, that the two animals are within the interaction region at time $t$ starting from the initial coordinates $\left(\vb*{r}^0_+,\vb*{r}^0_-\right)$,
\begin{equation}
\mu(t)=\int \dd{\vb*{r}_+} \int^{\prime} \dd{\vb*{r}_-} \, \Pi\left(\vb*{r}_+,\vb*{r}_+^0,\vb*{r}_-,\vb*{r}_-^0,t\right),
\label{eq:mut}
\end{equation}
where $\Pi\left(\vb*{r}_+,\vb*{r}_+^{0},\vb*{r}_-,\vb*{r}_-^{0},t\right)$ is the joint occupation probability solution of Eq. (\ref{eq:2walkOU}), when $\mathcal{C}=0$, given the initial conditions $\vb*{r}_+^{0}$ and $\vb*{r}_-^{0}$, referred to as the propagator (solution). It is simply given by the product of two-dimensional Ornstein-Uhlenbeck propagators for each animal centred around their respective focal point or home range centre \cite{sugayakenkre2018}. While $\mu(t)$ depends on the animal initial conditions (to lighten the formalism we have omitted this aspect from the notation),
the function $\nu(t)$ does not have any spatial dependence and is the probability, in the absence of any interaction, that the locations of the two animals are within the interaction region at a time $t$ after starting within it,
\begin{equation}
\nu(t)=\frac{\int \dd{\vb*{r}_+} \int^{\prime} \dd{\vb*{r}_-^{\prime}} \int^{\prime} \dd{\vb*{r}_-} \, \Pi\left(\vb*{r}_+,\vb*{r}_+^{\prime},\vb*{r}_-,\vb*{r}_-^{\prime},t\right)}{\int^{\prime} \dd{\vb*{r}_-^{\prime}}}.
\label{eq:nut}
\end{equation}

Note that while $\mathcal{T}_{\vb*{r}^0_{\pm}}(t)$ is normalised in time and has units of inverse of time, $\nu(t)$ and $\mu(t)$ are dimensionless quantities, but are not normalised in time, thus are not time probability density per se. One may notice in fact that, since $\Pi\left(\vb*{r}_+,\vb*{r}_+^{0},\vb*{r}_-,\vb*{r}_-^{0},t\right)$ is normalised in space, by integrating Eq. (\ref{eq:mut}) over all relative distance values, $\bm{r}_-$, $\mu(t)$ would equal exactly 1. This mathematical remark is equivalent to stating that, in the absence of interactions, there is certainty that the two animals are somewhere in space.

\subsection{The discrete space-time formalism}
\label{sec:discrete}

The recent development of the discrete space-time approach follows in the footsteps of the original studies on exciton annihilation in unbounded and periodic lattices \cite{kenkre_exciton_1982} and has extended that formalism to bounded domains with reflecting boundaries \cite{giuggiolisarvaharman2022}, to scenarios when the movement is altered by random resetting to a given location \cite{dasgiuggioli2022}, to dynamics in presence of spatial heterogeneities such as global biases \cite{sarvaharmangiuggioli2020}, variable diffusivities in space \cite{sarvaharmangiuggioli2023}, permeable barriers \cite{kaygiuggioli2022,sarvaharmangiuggioli2023} and different media and interfaces \cite{dasgiuggioli2023}.

The equation governing the transmission problem between two lattice random walkers is similar to the continuous version, but with the notable difference that the dynamics for the informed and uninformed individuals are governed by a difference equation rather than an integro-differential equation as in (\ref{eq:2walkOU}). By calling $\mathcal{P}(\bm{n}_1,\bm{n}_2,t)$ the joint occupation probability at discrete time $t$ for one walker to be at site $\bm{n}_1$ and the other at site $\bm{n}_2$, one has
\begin{align}
\mathcal{P}(\bm{n}_1,\bm{n}_2,t+1)=\sum_{\bm{\ell}_1,\bm{\ell}_2}\left[\mathbb{A}(\bm{n}_1,\bm{\ell}_1,\bm{n}_2,\bm{\ell}_2)\mathcal{P}(\bm{\ell}_1,\bm{\ell}_2,t)-\rho \sideset{}{'}\sum_{\bm{s}}\,\delta_{\bm{n}_1,\bm{s}}\,\delta_{\bm{n}_2,\bm{s}}\,\mathbb{A}(\bm{s},\bm{\ell}_1,\bm{s},\bm{\ell}_2)\mathcal{P}(\bm{\ell}_1,\bm{\ell}_2,t)\right].
\label{eqn:P_multi_defect}
\end{align}
In Eq. (\ref{eqn:P_multi_defect}) the elements of the tensor $\mathbb{A}(\bm{w}_1,\bm{\omega}_1,\bm{w}_2,\bm{\omega}_2)$ represent the transition probabilities at each time step for the first walker to move from site $\bm{\omega}_1$ to site $\bm{w}_1$ and for the second walker to move from site $\bm{\omega}_2$ to site $\bm{w}_2$. As we consider that the two individuals move independently of one another, $\mathbb{A}=\mathbb{B}_1\otimes \mathbb{B}_2$ where $\mathbb{B}_1(\bm{w}_1,\bm{\omega}_1)$ and $\mathbb{B}_2(\bm{w}_2,\bm{\omega}_2)$ control, respectively, the movement steps of walker 1 and walker 2. Compared to the continuous case, the interaction term in (\ref{eqn:P_multi_defect}) is now a summation rather than an integral, with $\delta_{\bm{a},\bm{b}}$ a Kronecker delta and the prime symbol indicating all lattice sites where interaction may occur, while $\rho$ represents the transfer probability once the two walkers are within the interaction range, and it is in place of the rate of the transfer $\mathcal{C}$ of the continuous case.
Note that while we use discrete time variables, it is straightforward to convert Eq.~(\ref{eqn:P_multi_defect}) to continuous time and changing accordingly jump probabilities to rates. There is however a computational convenience in using discrete versus continuous time in our context, and that is in the ease to invert to discrete time a generating function (i.e. a discrete Laplace transform) as compared to inverting to continuous time a function defined in the Laplace domain~\cite{giuggioli_exact_2020}.

One of the advantage of the spatially discrete formalism over the spatially continuous one is that it allows to quantify analytically the so-called splitting probability of interaction, that is the (time-dependent) joint probability that a transmission event occurs in a set of lattice sites or nodes and not in others. This prescription is naturally constructed in discrete space given the ease with which to associate the joint presence or absence of individuals at a set of $M$ locations with coordinates $\bm{S}_m=(\bm{s}_m,\bm{s}_m)$ ($m=1,...,M$), where the first and second coordinates refer, respectively, to the first and second animal. Given the (unordered) set $\bm{S}_m$ where the two individuals may transfer information, the probability that a transmission event (in any of the possible locations) occurs at time $t$ for the first time (first-transmission probability) is given by \cite{giuggiolisarvaharman2022}
\begin{align}
\mathcal{T}_{\bm{n}_0}(t)=\sum_{m=1}^M\mathcal{T}^{(m)}_{\bm{n}_0}(\rho,t) ,
\label{eq:int_prob}
\end{align}
where $\bm{n}_0=(\bm{n}_{1_0},\bm{n}_{2_0})$ represents the initial location of the two animals, and
$\mathcal{T}^{(m)}_{\bm{n}_0}(\rho,t)$ is the time-dependent probability that the transmission event occurs when the animals are at the lattice coordinates $\bm{S}_m$ and not at any of the other $M-1$ sites of interaction, the so-called splitting probabilities.

If $\Psi_{\bm{n}_{1_0},\bm{n}_{2_0}}(\bm{n}_1,\bm{n}_2,t)$ is the propagator of Eq. (\ref{eqn:P_multi_defect}) in the absence of any interaction ($\rho=0$), one can write the generating function---$\widetilde{f}(z)=\sum_{t=0}^{\infty}z^t f(t)$ for a generic function  $f(t)$---of the splitting probabilities, i.e. $\widetilde{\mathcal{T}}^{(m)}_{\bm{n}_0}(\rho,z)=\sum_{t=0}^{\infty}z^t\mathcal{T}^{(m)}_{\bm{n}_0}(\rho,t)$ as the following ratio \cite{giuggiolisarvaharman2022}
\begin{align}
\widetilde{\mathcal{T}}^{(m)}_{\bm{n}_0}(\rho,z)= \frac{\det[\mathbb{S}^{(m)}(\rho,z)]}{\det[\mathbb{S}(\rho,z)]},
\label{eq:split}
\end{align}
with $\mathbb{S}_{ii}(\rho,z)=(1-\rho)/\rho +\widetilde{\Psi}_{\bm{s}_{i},\bm{s}_{i}}(\bm{s}_{i},\bm{s}_{i},z)$ and $\mathbb{S}_{ij}(\rho,z)=\widetilde{\Psi}_{\bm{s}_{j},\bm{s}_{j}}(\bm{s}_{i},\bm{s}_{i},z)$, and $\mathbb{S}^{(m)}(\rho,z)$ the same as $\mathbb{S}(\rho,z)$ but with the vector $(\widetilde{\Psi}_{\bm{n}_{1_0},\bm{n}_{2_0}}(\bm{s}_{1},\bm{s}_{1},z), \widetilde{\Psi}_{\bm{n}_{1_0},\bm{n}_{2_0}}(\bm{s}_{2},\bm{s}_{2},z),...,\widetilde{\Psi}_{\bm{n}_{1_0},\bm{n}_{2_0}}(\bm{s}_{M},\bm{s}_{M},z))^T$ replacing the $m$-th column (the symbol $T$ indicates transpose). Note that $\Psi_{\bm{n}_{1_0},\bm{n}_{2_0}}(\bm{n}_1,\bm{n}_2,t)$ is the discrete analogue of the joint occupation probability used in the continuous variable section, which was expressed in terms of the transformed variable $\left(\vb*{r}_+,\vb*{r}_-\right)$.

To represent animals roaming within their own home ranges we consider two cases. In the first, the home ranges have partial overlap and the range where animals move is bounded by impenetrable boundaries (reflected random walkers). In the second, the domain is periodic, but large enough to be effectively unbounded, and each animal resets at random times to its own focal point (resetting random walkers). In both cases we consider the individuals to move independently, leading to a product form of the propagator for the process without transmission ($\rho=0$), namely
$\Psi_{\bm{n}_{1_0},\bm{n}_{2_0}}(\bm{n}_1,\bm{n}_2,t)=Q_{\bm{n}_{1_0}}(\bm{n}_1,t) Q_{\bm{n}_{2_0}}(\bm{n}_2,t)$ where $Q_{\bm{n}_{0}}(\bm{n},t)$ are the occupation probabilities for each independent walker.

For computational convenience we consider that an interaction event may occur only when the animals are co-located and we study both the one and two-dimensional scenarios. For the one-dimensional case we focus on the first-encounter probability, that is we set $\rho=1$, and we compute, through a numerical inversion of the generating function, the time-dependence of the first-encounter probability, offering a quantitative comparison with the corresponding discrete equivalent of $\mu(t)$ in Eq. (\ref{eq:mu_t}), which is given by
\begin{align}
    \mu(t) = \sum_{m=1}^{M}  \Psi_{\bm{n}_{1_0},\bm{n}_{2_0}}(\bm{s}_m,\bm{s}_m,t).
    \label{eq:mut-discrete}
\end{align}
Note that also in this discrete case $\mu(t)$ could be rewritten in terms of relative coordinates, but since we are considering only co-locations as encounters, it has no advantage.
For the two-dimensional case we limit ourselves to the analysis of the mean first-transmission time with reflected random walkers, but no comparison can be made to a corresponding mean for $\mu(t)$ given that it is not a normalised probability function and the evaluation of an average, via $\sum_{t=0}^{\infty} t \mu(t)$, is not finite.

\subsubsection{Diffusion in partially overlapping range-limited one-dimensional domains}

We consider that each animal diffuses within its own one-dimensional lattice domain, both of size $N$, and that the two domains overlap only partially. In this case the tensors $\mathbb{B}_m$ ($m=1$ and 2) reduce to matrices and their elements are $\mathbb{B}_{m_{ii}}=1-q_m$, $\mathbb{B}_{m_{ij}}=\mathbb{B}_{m_{ji}}=(\delta_{i,j+1}+\delta_{i,j-1})q_m/2$ when away from the boundary sites and $\mathbb{B}_{m_{11}}=\mathbb{B}_{m_{NN}}=1-q_m/2$. The actual dimension of the overlap region, that is the number of lattice sites $M$ where the animals may transmit information or encounter one another, is directly related to the distance $H=|c_1-c_2|$ between the central locations of the home ranges $c_1$ and $c_2$ via $M=N-H$. The individual walker propagator in this case is given by \cite{giuggioli_exact_2020}
\begin{align}
Q_{n_0}({n},t) = \sum_{k = 0}^{N - 1} h^{(N)}_{k} (n, n_0) \Big[ 1+ s^{(N)}_{k} \Big]^t,
\label{eq:PsiNR-gen_time}
\end{align} 
where
\begin{align}
h^{(N)}_{k} (n, n_0)=\frac{\alpha_k}{N}\cos\left[\left(n-\frac{1}{2}\right)\frac{\pi k}{N}\right]\cos\left[\left(n_0-\frac{1}{2}\right)\frac{\pi k}{N}\right]
\label{eq:hk}
\end{align}
with $\alpha_0=1$ and $\alpha_k=2$ for $k\geq 1$, and
\begin{align}
s^{(N)}_k=q\left[\cos\left(\frac{\pi k}{N}\right)-1\right],
\label{eq:sk}
\end{align}
for the first animal, and $h^{(N)}_{k} (n-H, n_0-H)$ with $n=1+H,...,N+H$ for the second animal.
From $\Psi_{n_{1_0},n_{2_0}}(n_1,n_2,t)=Q_{n_{1_0}}(n_1,t) Q_{n_{2_0}}(n_2,t)$, it is straightforward to obtain the generating function $\widetilde{\Psi}_{n_{1_0},n_{2_0}}(n_1,n_2,z)$, and use it to construct $\widetilde{\mathcal{T}}_{\bm{n}_0}(z)$.

\subsubsection{Diffusion with resetting in one-dimensional domains}

For the case of the resetting random walkers, one requires to modify Eq. (\ref{eqn:P_multi_defect}) by adding on the right hand side the terms $r_1\delta_{n_1,c_1}$ and $r_2\delta_{n_2,c_2}$, with $r_1$ and $r_2$ representing the probability for the first and second walker to relocate at random times to site $c_1$ and $c_2$,  respectively. In this case, the tridiagonal matrices are given by $\mathbb{B}_{m_{ij}}=\mathbb{B}_{m_{ji}}=(\delta_{i,j+1}+\delta_{i,j-1})(1-r_m)q_m/2$ and $\mathbb{B}_{m_{ii}}=(1-r_m)(1-q_m)$.
By taking periodic boundary conditions, the propagator for an individual resetting random walker is given by \cite{dasgiuggioli2022}
\begin{align}
Q_{n_{0}}(n,t )  &=  r \sum_{k = 0}^{N - 1}  g^{(N)}_{k} (n, c)  ~\frac{\gamma_{k}^{t}- 1}{\gamma_k-1} +   \sum_{k = 0}^{N - 1} g^{(N)}_{k} (n, n_{0})  ~\gamma_{k}^{t} \, , \label{eq:QReset-bounded_per} 
\end{align}
where $c$ is the resetting site, $\gamma_k=(1-r)\left[1+s^{(N)}_k\right]$ with $s_k$ given in Eq. (\ref{eq:sk}) and
$g^{(N)}_{k} (x,y)=\cos[2\pi k(x-y)/N]/N $. Analogously to the reflecting case above, the propagator for both walkers, that is the solution of Eq. (\ref{eqn:P_multi_defect}) in the absence of transmission events, is given by $\Psi_{n_{1_0},n_{2_0}}(n_1,n_2,t) = Q_{n_{1_0}}(n_1,t) Q_{n_{2_0}}(n_2,t)$.

\subsubsection{Diffusion in two-dimensional range-limited home ranges}

For a two-dimensional setting we consider animals living in home ranges of rectangular shape of identical size. The range limitation of the animals is ensured by reflecting boundary conditions. The two home ranges are aligned along the vertical axis, but are shifted by an amount equal to $H$ sites along the horizontal axis. In the absence of interactions, for each animal the propagator is given by \cite{giuggioli_exact_2020}
\begin{align}
Q_{\bm{n}_{0}}(\bm{n},t) =\sum_{k=0}^{N-1}\sum_{\ell=0}^{\mathcal{N}-1}h^{(N)}_k(n_x,n_{x_0})h^{(\mathcal{N})}_{\ell}(n_y,n_{y_0})\left[1+\frac{s^{(N)}_k}{2}+\frac{s^{(\mathcal{N})}_{\ell}}{2}\right]^t,
\label{eq:Q_2d_refl}
\end{align}
where $N$ and $\mathcal{N}$ represent, respectively, the number of sites along the two directions and with $h^{(L)}_{\omega}(n,m)$ and $s^{(L)}_{\omega}$ given, respectively, in Eqs. (\ref{eq:hk}) and (\ref{eq:sk}).

To determine the mean-transmission time at any of the co-locations one requires knowledge of the mean first-passage time between the initial location and the co-location sites, the mean first-passage between all co-location sites (all permutations), and the mean return time to the co-location sites. For that we use Eq. (\ref{eq:Q_2d_refl}) to build the product of the individual propagators in time by shifting by $H$ sites the coordinates of the horizontal axis for the second individual. For an initial condition with coordinates $\bm{n}_0=(n_{x_0},n_{y_0})$ along the horizontal and vertical axes and with $\bm{n}=(n_x,n_y)$, we construct the generating function of the 4-dimensional propagator,,
\begin{align}
\widetilde{\Psi}_{\bm{n}_{1_0},\bm{n}_{2_0}}(\bm{n}_1,\bm{n}_2,z)&=\sum_{t=0}^{\infty}z^t Q_{\bm{n}_{1_0}}(\bm{n}_1,t)Q_{\bm{n}_{2_0}}(\bm{n}_2,t)\nonumber \\
&=\sum_{k_1=0}^{N-1}\sum_{\ell_1=0}^{\mathcal{N}-1}\sum_{k_2=0}^{N-1}\sum_{\ell_2=0}^{\mathcal{N}-1}\frac{h^{(N)}_{k_1}(n_{1x},n_{{1x}_0})h^{(\mathcal{N})}_{\ell_1}(n_{1y},n_{{1y}_0})h^{(N)}_{k_2}(n_{2x}-H,n_{{2x}_0}-H)h^{(\mathcal{N})}_{\ell_2}(n_{2y},n_{{2y}_0})}{1-z\Big[1+\frac{s^{(N)}_{k_1}}{2}+\frac{s^{(\mathcal{N})}_{\ell_1}}{2}\Big]\Big[1+\frac{s^{(N)}_{k_2}}{2}+\frac{s^{(\mathcal{N})}_{\ell_2}}{2}\Big]},
\label{eq:prop_4d}
\end{align}
with the range in $n_{1x}$ and $n_{2x}$ being, respectively, $[1,N]$ and $[1+H,N+H]$, while the range for both  $n_{1y}$ and $n_{2y}$ is $[1,\mathcal{N}]$.
From Eq. (\ref{eq:prop_4d}) it is straightforward to obtain the mean (first) return time \cite{kac_notion_1947} to a site $\bm{n}=(n_{1x},n_{1y},n_{2x},n_{2y})$,
\begin{align}
\mathcal{R}_{\bm{n}}=\Bigg[h^{(N)}_{0}(n_{1x},n_{1x})h^{(\mathcal{N})}_{0}(n_{1y},n_{1y})h^{(N)}_{0}(n_{2x}-H,n_{2x}-H)h^{(\mathcal{N})}_{0}(n_{2y},n_{2y})\Bigg]^{-1},
\label{eq:mrt}
\end{align}
and through a simple differentiation \cite{rednerbook2001}, i.e. $T_{(\bm{n}_{1_0},\bm{n}_{2_0})\rightarrow (\bm{n}_1,\bm{n}_2)}=\left.\frac{\mathrm{d}}{\mathrm{d}z} \left[\frac{\widetilde{\Psi}_{\bm{n}_{1_0},\bm{n}_{2_0}}(\bm{n}_1,\bm{n}_2,z)}{\widetilde{\Psi}_{\bm{n}_{1},\bm{n}_{2}}(\bm{n}_1,\bm{n}_2,z)}\right]\right|_{z=1}$, the mean first-passage time
\begin{align}
&\hskip-20pt T_{(\bm{n}_{1_0},\bm{n}_{2_0}) \rightarrow (\bm{n}_1,\bm{n}_2)} \nonumber \\
&=2\underset{k_1+k_2+\ell_1+\ell_2>0}{\sum_{k_1=0}^{N-1}\sum_{\ell_1=0}^{\mathcal{N}-1}\sum_{k_2=0}^{N-1}\sum_{\ell_2=0}^{\mathcal{N}-1}}\Bigg[h^{(N)}_{k_1}(n_{1x},n_{{1x}_0})h^{(\mathcal{N})}_{\ell_1}(n_{1y},n_{{1y}_0})h^{(N)}_{k_2}(n_{2x}-H,n_{{2x}_0}-H)h^{(\mathcal{N})}_{\ell_2}(n_{2y},n_{{2y}_0})\Bigg. \nonumber \\
&\Bigg.
-h^{(N)}_{k_1}(n_{1x},n_{1x})h^{(\mathcal{N})}_{\ell_1}(n_{1y},n_{1y})h^{(N)}_{k_2}(n_{2x}-H,n_{2x}-H)h^{(\mathcal{N})}_{\ell_2}(n_{2y},n_{2y})\Bigg]\Bigg\{h^{(N)}_{0}(n_{1x},n_{1x})h^{(\mathcal{N})}_{0}(n_{1y},n_{1y})\Bigg.\nonumber \\
&\Bigg.\times h^{(N)}_{0}(n_{2x},n_{2x})h^{(\mathcal{N})}_{0}(n_{2y},n_{2y})\Big[\left(s^{(N)}_{k_1}+s^{(\mathcal{N})}_{\ell_1}\right)\left(s^{(N)}_{k_2}+s^{(\mathcal{N})}_{\ell_2}\right)+s^{(N)}_{k_1}+s^{(\mathcal{N})}_{\ell_1}+s^{(N)}_{k_2}+s^{(\mathcal{N})}_{\ell_2}\Big]\Bigg\}^{-1},
\label{eq:mean_transm}
\end{align}
between a starting site $\bm{n}_0=(n_{{1x}_0},n_{{1y}_0},n_{{2x}_0},n_{{2y}_0})$ and a target site $\bm{n}=(n_{1x},n_{1y},n_{2x},n_{2y})$.

\section{Results}

\subsection{Difference between first-encounter probability and distance threshold probability}
\label{sec:diff_cont}

We consider the case of perfect transfer efficiency, $\mathcal{C}\rightarrow \infty$ in Eq. (\ref{eq:inf_prob_lap}), and focus on the so-called hard encounter events, that is those instances when animals reach a relative distance $b$. By integrating over all possible angles and separation up to radius $b$ in Eqs. (\ref{eq:mut}) and (\ref{eq:nut}), following Kenkre and Sugaya \cite{sugayakenkre2018}, one obtains
\begin{align}
\mu(t)=1-Q_1\left(\frac{\mathcal{F}(r^0,\phi^0,t)}{\sqrt{4Dh(t)}},\frac{b}{\sqrt{4Dh(t)}}\right)
\label{eq:mu_t}
\end{align}
and
\begin{align}
\nu(t)=1- \frac{1}{\pi b^2} \int_{0}^{b} \dd{r'} r' \int_{0}^{2 \pi} \dd{\phi'} Q_1 \left( \frac{ \mathcal{F}(r',\phi',t)}{\sqrt{4D h(t)}} ,  \frac{ b }{\sqrt{4D h(t)}}\right),
\label{eq:nu_t}
\end{align}
where $\mathcal{F}^2(r,\phi,t)=2 r H \cos(\phi - \omega) - H^2 - \ee^{-2 \gamma t} \big[ 2 r H \cos(\phi - \omega) - H^2-r^2 \big]$, with $H$ and $\omega$, respectively, the distance and relative angle between the home range centres, where $h(t)=\left[1-\exp(-2\gamma t)\right](2\gamma)^{-1}$, and where  $Q_{1}(s_1,s_2)=1-\int_{0}^{s_2}dz\,z\exp\left[-\left(z^2+s_1^2\right)/2\right]I_0\left(s_1z\right)$, is the Marcum $Q$-function of order 1. Given that $\mu(t)$ is a spatial integration of the (time-dependent) joint occupation probability over the relative distance $b$, we refer to it in the following as the distance threshold probability.

Equation (\ref{eq:mu_t}), with a rate constant multiplying it, has been called the mean encounter rate (Eq. (14) in Ref. \cite{martinez-garcia_how_2020} and has been proposed to explore how the interplay between the scale of perception and home-range size affect encounter rates. Although the discrepancy with Eq. (\ref{eq:inf_prob_lap}) when $\mathcal{C}\rightarrow \infty$ is self-evident, it is instructive to rewrite Eq. (\ref{eq:inf_prob_lap}) in that limit as $\widetilde{\mathcal{E}}(\epsilon)\widetilde{\nu}(\epsilon)=\widetilde{\mu}(\epsilon)$, renaming first-transmission as first-encounter, $\mathcal{T}(t)\overset{\mathcal{C}\rightarrow \infty}{\rightarrow}\mathcal{E}(t)$, and through a Laplace inversion obtain
\begin{align}
\mu(t)=\int_0^t \dd{s} \mathcal{E}(t-s)\nu (s).
\label{eq:mu_t_2}
\end{align}
Equation (\ref{eq:mu_t_2}) shows the relation between the first-encounter probability, $\mathcal{E}(t)$, and the distance threshold probability, and its structure is quite revealing. It represents a generalisation of the well-known renewal equation for Markov processes \cite{rednerbook2001}, $P_{x_0}(x,t)=\int_0^t \dd{s}  F_{x_0\rightarrow x}(t-s)P_x(x,s)$, that relates the occupation probability $P_{x_0}(x,t)$ to be at $x$ at time $t$ starting at $x_0$ with the first-passage or first-hitting probability, $F_{x_0\rightarrow x}(t)$ to reach $x$ from $x_0$. While it may seem always possible to write an equation such as (\ref{eq:mu_t_2}), with $\mu(t)$ and $\nu(t)$ representing a spatially integrated version, or more precisely integration over a given range, of $P_{x_0}(x,t)$ and $P_x(x,t)$, respectively, it turns out to be true only when 
$\int \dd{\vb*{r}_+} \int^{\prime} \dd{\vb*{r}_-} \Pi\left(\vb*{r}_+,\vb*{r}_+^{\prime},\vb*{r}_-,\vb*{r}_-^{\prime},t\right)$ is independent of $\vb*{r}_+^{\prime}$, something that occurs only when certain spatial symmetries are present. While it is difficult to visualise the geometry of these special cases with animals moving in two and three dimensions, given that the set of locations where encounters may occur are part of a 4 or 6-dimensional space, it may help to think about a one-dimensional encounter problem. The simplest scenario is that of two Brownian walkers that diffuse without any spatial constraint on a line and come into `contact' once they are at a distance $b$. Their encounter dynamics can be mapped onto the search dynamics of a two-dimensional Brownian walker that hits for the first time a radial target of radius $b$ centred around the origin. The associated Eq. (\ref{eq:mu_t_2}) becomes equivalent to an effective one-dimensional renewal equation since a first hitting event is controlled only by the radial coordinate of the Brownian walker being equal to $b$. More intuitively, whenever a set of interaction locations are arranged spatially as a single big target, then one may potentially write equations such as (\ref{eq:inf_prob_lap}) and (\ref{eq:mu_t_2}) where $\mu(t)$ and $\nu(t)$ are spatially integrated representation of the animals' occupation probability in the absence of any interaction.

More generally, in all scenarios that lack high spatial symmetries, the interaction locations have a complicated geometry and parametrising the resulting shape with multiple variables becomes a
complicated task. In addition, when a first-hitting event requires to specify the threshold value of many variables, one needs to construct splitting probabilities, practically separating the space into multiple areas. In these situations identifying these separate areas where interactions may occur is easily met by mapping the dynamics into discrete space and studying the first-transmission to a set of multiple targets on a lattice, which is the subject of the next subsections.

\subsection{First-encounter probability with overlapping home ranges in one dimension}

\begin{figure}[!htbp]
\centering
\includegraphics[scale=1.]{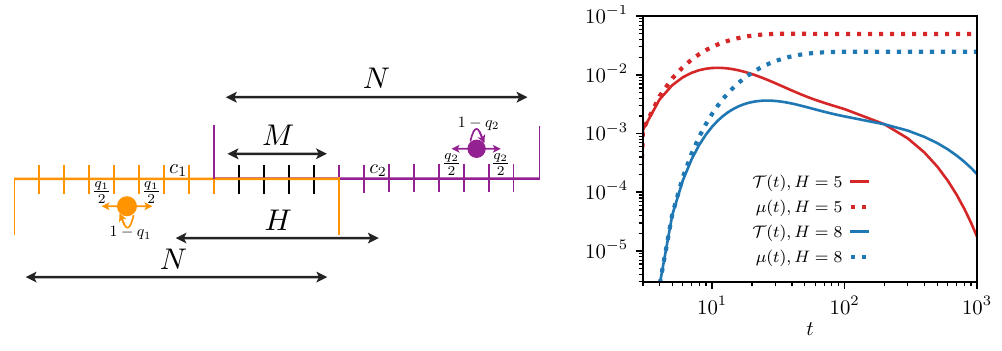}
\caption{Schematics of two animals roaming within separate one-dimensional home ranges with partial overlap (left panel) and their first-encounter probability (right panel). On the left panel the circle displays an animal while the arrows indicate the movement probability at each time step: the left and right horizontal arrows represent the probability to move, respectively, left and right, while the bent arrow is the probability of remaining at the same site. Although not shown in the schematics, the movement rules at the boundary sites are slightly different with the probability of staying modified to $1-q_m/2$, while the probability to move outside of the domain is suppressed. The size of the two home ranges is equal to $N=11$. The first walker diffuses within a domain centred at $c_1=6$, is limited by reflecting boundaries at sites 1 and 11, and it starts from $n_{1_{0}}=6$, while for the second walker there are two cases: the allowed range is either (i) [6,16] or (ii) [9,19], and in both cases with reflecting boundaries at the end sites. The two animals may encounter each other when they simultaneously occupy a site in the overlap region, made up of a total of $M$ sites. 
In case (i), the distance between the two home range centres is $H=5$ and the second walker starts from $n_{2_0} = 11$, while in case 
(ii), we have  $H=8$ and $n_{2_0} = 14$. 
The quantities $\mathcal{T}_{\bm{n}_0}(t)$ (in the legend we have omitted the subscript $\bm{n}_0$ for clarity), from Eq.~\eqref{eq:int_prob}, and $\mu(t)$, from Eq.~\eqref{eq:mut-discrete}, are shown by the continuous and dotted lines (in red for case (i) and in blue for case (ii)), respectively. 
For both walkers, we take diffusivity $q_1=q_2=0.4$.}
\label{fig:encounter-refl-box}
\end{figure}

Having shown formally in the example studied in Sec. \ref{sec:diff_cont} the relation between $\mathcal{E}(t)$ and $\mu(t)$, we now proceed to quantify their difference with the discrete formalism. For simplicity and computational convenience we start by considering animals living in one dimensional domain bounded by reflecting boundaries, as depicted in the left panel of Fig. \ref{fig:encounter-refl-box}. Past analyses to determine the transmission dynamics in this one-dimensional problem has lead to analytic expressions only for the mean transmission time \cite{giuggiolietal2013}, whereas we are now able to capture the exact dynamics for the entire transmission probability $\mathcal{T}_{\bm{n}_0}(t)$. We consider two different home range overlaps with the two animals starting, respectively, at $c_1$ and $c_2$, and use
standard inversion routines (i.e. a one dimensional integration) for generating functions \cite{abate_numerical_1992,abateetal2000} to plot the first-encounter probability in the right panel of Fig. \ref{fig:encounter-refl-box}.

As a comparison we plot the discrete analogue of the function $\mu(t)$, namely Eq. (\ref{eq:mut-discrete}).
While $\mathcal{T}_{\bm{n}_0}(t)$ decays to zero at long times, $\mu(t)$ reaches a finite non-zero value, making it evident why the former is a normalised probability function, while the latter is not. The long time saturation value of $\mu(t)$ indicates that once the memory of the initial placement vanishes the chance that two individuals are found in any of the possible co-locations is constant and equals the integral over the interaction region of the steady state joint occupation probability. 

\subsection{First-encounter probability with one dimensional resetting dynamics}

We take the so-called resetting random walker as another representation of an animal that moves within a  home range. As the walker resets at random times to a focal point in space, the range of movement is effectively bounded, with the resetting sites representing the den or burrow where animals tend to return to. At long times the spatial occupation probability is in fact equivalent to a steady state probability if the waker were to move with a constant bias towards the resetting location \cite{giuggioli2019comparison}. For computational convenience we take a periodic spatial domain for both walkers.
Even though the movement of the walkers is effectively bounded, and differently from the reflected walker case above, we need to specify a finite number $M$ of interacting locations given that the discrete formalism requires to evaluate a determinant of size $M$. With the appropriate choice of the movement model parameters and the placement and number of interacting locations around the home range centres, we ensure that the probability of transmission at the sites excluded from the $M$ selected is negligible.

\begin{figure}[!htbp]
\centering
\includegraphics[scale=1.]{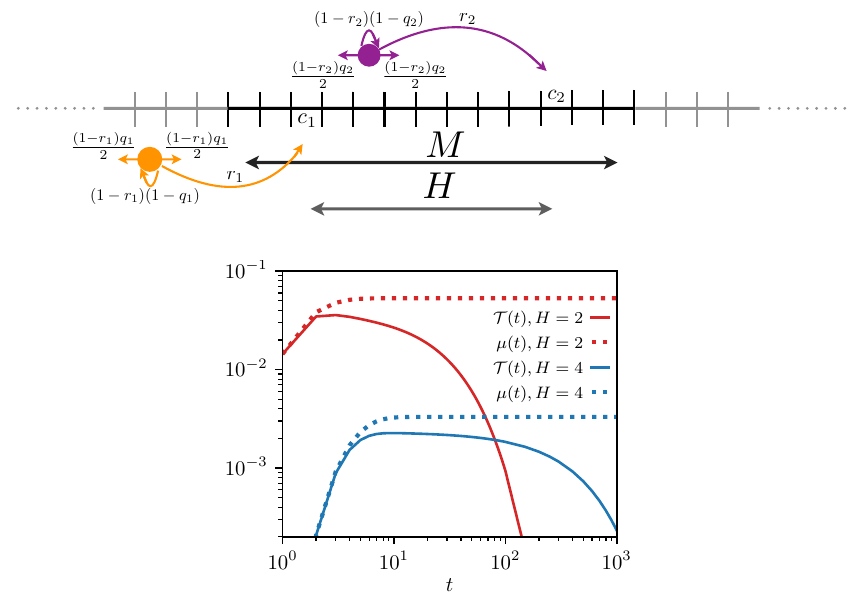}
\caption{Schematics of two resetting random walkers (top panel) and their first-encounter probability (bottom panel). Compared to Fig. \ref{fig:encounter-refl-box}, the movement rules are modified by the fact that at each time step the animal may reset its location to its own home range centre, indicated in the schematics by the long arrows with probability $r_1$ and $r_2$. To mimic unbounded space the boundary conditions are periodic and the domain size ($N=19$), diffusivity ($q_1=q_2=0.4$) and resetting probability ($r_1=r_2=0.4$) are chosen to ensure that the contributions to the encounters of those trajectories that exploit the lattice periodicity are negligible. For two cases analysed the home range centres are located at $(c_1,c_2)=(9,11)$ and $(c_1,c_2)=(8,12)$, giving, respectively, a relative distance $H$ between their home range centre of 2 and 4.
We have used Eq. (\ref{eq:QReset-bounded_per}) to construct $\mathcal{T}_{\bm{n}_0}(t)$ in Eq.~\eqref{eq:int_prob} and $\mu(t)$ in Eq.~\eqref{eq:mut-discrete}, and display them with the continuous and dotted lines (in red for case (i) and in blue for case (ii)), respectively. 
In both cases, the walkers start from their corresponding home range centres ($c_1,c_2$) and interact when they simultaneously occupy a site within the domain [7,13], consisting of $M=7$ sites.}
\label{fig:encounter-reset}
\end{figure}

We consider perfect transfer efficiency and compare the first-encounter probability, $\mathcal{T}_{\bm{n}_0}(t)$ with $\rho=1$, to $\mu(t)$ in Fig. \ref{fig:encounter-reset}. Compared to the previous case with reflecting walkers, one can see that the dynamics is relatively quicker. The first-encounter mode is reached after ten and twelve steps when, respectively, $H=5$ and $H=8$ in Fig. \ref{fig:encounter-refl-box}, while it is reached after three steps when $H=2$ and after eight steps when $H=4$ in Fig. \ref{fig:encounter-reset}.
This faster time dependence can be explained by the choice of the
parameters of the problem. In the resetting case at each time step the chance of a walker to move can be shown to be 3/10 relative to the reflecting walkers.
This fast dynamics is also noticeable in $\mu(t)$, when compared to Fig. \ref{fig:encounter-refl-box}.

\subsection{Mean first-transmission times between animals diffusing in two-dimensional home ranges}

\begin{figure}[!htbp]
\centering
\includegraphics[scale=1]{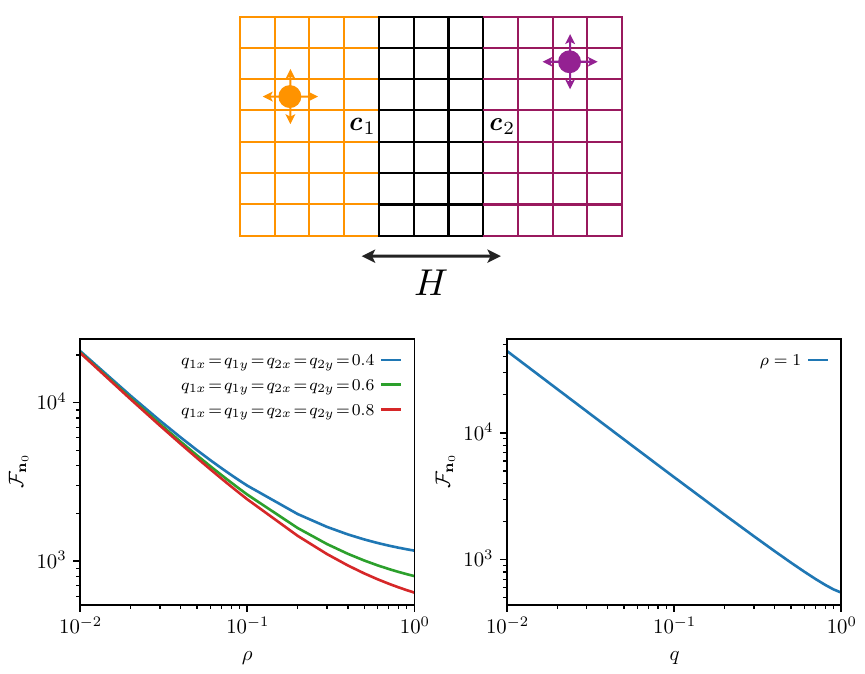}
\caption{Schematics of two animals roaming within separate two-dimensional home ranges with partial overlap along one direction (top panel) and their mean first-transmission time (bottom panel). On the top panel the circle displays an animal while the arrows indicate the movement probability at each time step: the left, right, up, and down arrows represent the probability to move, respectively, left, right, up, and down. 
Although not shown by an arrow, the $i$th animal while not at any of the boundaries can stay at the same site with probability $1-q_{ix}/2-q_{iy}/2$, where $q_{ix}$ and $q_{iy}$ denote the diffusivities in the $x$- and $y$-directions, respectively.
The probability of staying at sites (except four corners) on boundaries along the $x$- and $y$-directions is $1-q_{ix}/4-q_{iy}/2$ and $1-q_{ix}/2-q_{iy}/4$, respectively, while at the four corners it is $1-q_{ix}/4-q_{iy}/4$. 
The size of the two home ranges is equal to $N \times  \mathcal{N}$ with $N=11$ and $\mathcal{N} = 5$. The first walker diffuses within a domain centred at $\bm{c}_1=(6,3)$, is limited by reflecting boundaries at sites 1 and 11 in $x$-direction and at sites 1 and 5 in $y$-direction, and it starts from $\bm{n}_{1_{0}}=(6,3)$. For the second walker, the allowed range is [9,19] in $x$-direction and [1,5] in $y$-direction, and in both cases with reflecting boundaries at the end sites. Hence, the domain for the second walker is centred at $\bm{c}_2=(14,3)$, which is also its starting point, i.e., $\bm{n}_{2_0} = (14,3)$. 
The distance between the two home range centres is $H= | \bm{c}_2 - \bm{c}_1| = 8$.
The two animals may encounter each other when they simultaneously occupy a site in the overlap region, made up of a total of $M$ sites.
The quantity $\mathcal{F}_{\bm{n}_0}$ from Eq.~\eqref{eq:mean_int_time} is shown on the panels at bottom. The bottom left panel shows $\mathcal{F}_{\bm{n}_0}$ as a function of $\rho$ for the same diffusivities for both walkers in both directions.
The bottom right panel shows the mean encounter time $\mathcal{F}_{\bm{n}_0}$ ($\rho=1$) as a function of diffusivity $q$ such that $q=q_{1x}=q_{1y}=q_{2x}=q_{2y}$.}
\label{fig:mean-transmission-time-refl-box}
\end{figure}

As mentioned earlier the reaction diffusion approach allows to map the first-transmission problem with transfer efficiency $\rho$ to a first-absorption problem with multiple static partially absorbing targets located at $\bm{S}_i$ ($i=1,...,M$) in a spatial domain of double the original dimensions. Since theoretically it is now possible to predict exactly the mean first-absorption time to any of a set of partially absorbing targets \cite{giuggiolisarvaharman2022}, we exploit here that advance for our transmission problem. We examine the case of two reflected lattice walkers moving in two dimensions in partially overlapping home ranges (see top panel of Fig. \ref{fig:mean-transmission-time-refl-box}). If we call $\mathcal{F}_{\bm{n}_0}$, the mean-transmission time to a set of $M$ co-location sites starting from a site $\bm{n}_0$, we have \cite{giuggiolisarvaharman2022}
\begin{align}
\mathcal{F}_{\bm{n}_0}=\frac{\det\left(\mathbb{T}_0\right)}{\det\left(\mathbb{T}_1\right)-\det\left(\mathbb{T}\right)},
\label{eq:mean_int_time}
\end{align}
where the elements of the matrix $\mathbb{T}$ are expressed exactly in terms of mean-first passage times $T$, mean return times $\mathcal{R}$ and the transfer efficiency $\rho$. More specifically we have $\mathbb{T}_{ij}=T_{\bm{S}_j\rightarrow \bm{S}_{i}}$ ($j,i=1,...,M$, with $i\not=j$), while the diagonal elements are given by $\mathbb{T}_{ii}=\frac{\rho-1}{\rho}\mathcal{R}_{\bm{S}_{i}}$ where $\mathcal{R}_{\bm{S}_{i}}$ is the mean return time to site $\bm{S}_{i}$. The other two matrices are obtained from $\mathbb{T}$ as follows: $\mathbb{T}_{0_{ij}}=\mathbb{T}_{ij}-T_{\bm{n}_0\rightarrow \bm{S}_i}$ and $\mathbb{T}_{1_{ij}}=\mathbb{T}_{ij}-1$.

We use Eqs. (\ref{eq:mrt}) and (\ref{eq:mean_transm}) to build the elements of the matrices in (\ref{eq:mean_int_time}) and in Fig. \ref{fig:mean-transmission-time-refl-box} we plot $\mathcal{F}_{\bm{n}_0}$, the mean-transmission time as a function of $\rho$ for different diffusion constant in the bottom left panel, and the mean-encounter time ($\rho=1$) as a function of the the diffusion constant, expressed via the (dimensionless) diffusivity parameters $q_{i_x}$ and $q_{i_y}$. As $\rho$ approaches 1, the dynamics become motion limited, because the slowest process, the time to reach the targets in this case, governs the time scale of the interaction. From Eq. (\ref{eq:mean_int_time}) one can extract a perturbation expansion in $1/\rho-1$ \cite{giuggiolisarvaharman2022}, and the shape of the slowing down in the decrease of $\mathcal{F}_{\bm{n}_0}$ in the left panel can be quantitatively explained as the first order correction to the zeroth order (motion limited) term. The plot in the right panel shows that the encounter rate is mainly linearly proportional to the animal diffusion constant \cite{benichou_first-passage_2014}. While such dependence is somewhat expected, what is unexpected is the very limited deviation from an inverse $q$ dependence of $\mathcal{F}_{\bm{n}_0}$, because one can show that for any element $T_{(\bm{n}_{1},\bm{n}_{2})\rightarrow (\bm{m}_1,\bm{m}_2)}=q^{-1}g(\bm{n}_{1},\bm{n}_{2},\bm{m}_{1},\bm{m}_{2},q)$. The right panel thus points to a negligible dependence of the function $g(\bm{n}_{1},\bm{n}_{2},\bm{m}_{1},\bm{m}_{2},q)$ on $q$.

\section{Summary and discussion}
\label{sec:concl}

The ability to track simultaneously with high resolution a large number of animals both in laboratory settings and in the field demands the development of modelling approaches to predict when, where and how animals interact. As some of the theoretical challenges to represent animal interactions have already been tackled in analysing physical and chemical systems, our aim here has been to make the movement ecology community profit from insights already gained in other fields. To do so we have open up the modelling literature from statistical physics, both past and present, on reaction diffusion processes and we have studied the transmission and encounter problem between two animals leaving within separate home ranges.

We have presented the mathematical details that allow to predict over time first-transmission and first-encounter probability both in continuous and discrete variables. With continuous variables we have considered two Brownian walkers that may interact with an information transfer rate $\mathcal{C}$ when within a threshold distance $b$ and have modelled their motion via a Ornstein-Uhlenbeck process. With discrete variables we have instead considered that interactions may occur with probability $\rho$ upon co-location and have taken reflected and resetting lattice random walkers to represent animals that roam within distinct home ranges.

With perfect transfer efficiency ($\mathcal{C}\rightarrow \infty$ or $\rho\rightarrow 1$), the interaction events reduce to encounter events. In this case, we have compared the continuous formulation to study first-encounter probability to the one proposed recently in the literature using a distance threshold probability and we have been able to derive a mathematical equation that connects the two quantities. To quantify the difference in the two probabilities we have used discrete variables and looked at the dynamics of two animals living in separate home ranges and moving and interacting on constrained one-dimensional lattices. That comparison allows to visualise why one is a normalised probability function with all finite moment, while the other is not normalised and possess infinite moments.

We recognise that the first-encounter probability and the distance threshold probability capture different aspects of the animal dynamics, and we thus believe that there should be scope for employing both, or either, especially in light of the various mechanisms with which animals may interact in an ecological setting. If an encounter event affects detectable characteristics of the animals, then clearly the first instance when that happens is the relevant observable. Examples include the transfer of an infectious pathogen or a parasite, a predator capturing a prey, or an animal passing knowledge about food sources by being observed or smelled by a nearby conspecific. In all these circumstances the first-transmission probability is a necessary tool to predict the dynamics based on the interplay between the transfer efficiency and the rate of movement. If, on the other hand, information transfer upon interaction is not  binary (presence/absence) or it is hard to detect, then knowledge of when animals are within a given distance becomes useful, as shown in the very recent developments \cite{noonanetal2021,alberyetal2021,yangetal2023,noonanetal2023} following Ref. \cite{martinez-garcia_how_2020}.

While we have focused here on destructive searches, this does not preclude the use of the reaction formalism in non-destructive studies, and more specifically the one with discrete space-time variables. In non-destructive scenarios, as the evaluation of the forager efficiency is based upon the cumulative encounter of targets, the quantity of interest becomes the (multiple) visitation statistics to any of the lattice sites where targets are located, coupled with a resetting of the walker to a neighbouring site upon a target capture. Such dynamics can be studied analytically with the discrete formalism, which has general validity for any Markov movement process and irrespective of the choice of spatial constraint or boundary conditions or the presence of spatial heterogeneities. It could be exploited to provide some useful insights to some of the ongoing debate about the efficiency of stochastic searches when targets gets replenished and walkers move as L\'{e}vy walkers \cite{viswanathanetal1999,benhamou2007,reynolds2008,levernieretal2020,buldyrevetal2021} and to explore the dependence on the density \cite{jamesetal2008}, boundary conditions \cite{buldyrevetal2001,jamesetal2010} and the spatial distribution of the resources \cite{humphriessims2014} without using time consuming stochastic simulations. As the discrete formalism allows to include any type of heterogeneities, it could also bring insights on the timely studies about species survival following habitat fragmentation and habitat loss as a function of the animal foraging statistics \cite{wosniacketal2014,niebuhretal2015}. It is also worth mentioning another advantage of the discrete spatial formalism in comparison to the diffusion equation. With the latter it is well known that one describes an ensemble of spatio-temporal trajectories that include (with some exponentially small probability) those that move infinitely fast from a localised initial condition. This limitation, on the other hand, is not present when using random walks on a lattice.

Despite the limitation of our Markov assumption, which considers the movement to be diffusive, extensions of encounter estimations to situations where the assumption about persistence is relaxed are possible. The effects of correlations in the movement steps, also called motion coherence, can be incorporated in a general reaction-motion formalism using the so called generalised master equation \cite{kenkre1973,kenkre1977}, which possesses a non-local memory kernel with one extreme (never decaying memory) reducing to a wave equation, that is to ballistic motion, and the other extreme to an infinitely fast decaying memory, that is diffusive motion. The intermediate situation, with an exponentially decaying memory, represents coherent motion at short times, and incoherent motion at long times, and was shown to be identical to the telegraphers' equation in one dimension \cite{kenkre1977}. In the context of exciton annihilation, an example of how motion coherence has been included using a generalised master equation can be found in Ref. \cite{gulenetal1988}.

Accounting for correlations in the discrete formalism is also possible and can be accomplished by representing a movement process with $\tau$ correlated steps as a vectorial Markov process with $\tau$ components (see e.g. \cite{ernst1988}). The formal difference from the cases analysed here consists of the need to deal with larger matrices since the set of $M$ interaction locations would become $\tau M$ possible interaction sites in the higher dimensional space.

Overall, while there is still much development to be done, an important contribution of our study is that using a reaction motion formalism it is possible to predict time-dependent first-transmission, and in the limit, first-encounter probability in terms of the animal movement statistics and the geometric constraints of the space.

\section*{Conflict of Interest Statement}
The authors declare that the research was conducted in the absence of any commercial or financial relationships that could be construed as a potential conflict of interest.

\section*{Author Contributions}
VMK, RN and LG developed the idea of the work, DD and LG developed the math and DD prepared the figures, LG wrote the initial draft with contributions from all, and all authors contributed to reviewing and editing.

\section*{Funding}
LG acknowledges funding from the Biotechnology and Biological Sciences Research Council (BBSRC) Grant No. BB/T012196/1, and the National Environment Research Council (NERC) Grant No. NE/W00545X/1. RN acknowledges funding from the Minerva Center for Movement Ecology, the Minerva Foundation, and the Adelina and Massimo Della Pergola Chair of Life Sciences. All authors would like to thank the Isaac Newton Institute for Mathematical Sciences for support and hospitality during the programme `Mathematics of Movement: an interdisciplinary approach to mutual challenges in animal ecology and cell biology', when part of the work on this paper was undertaken, supported by the EPSRC Grant Number EP/R014604/1.

\bibliographystyle{unsrt}  
\bibliography{library}

\end{document}